\documentclass{appolb}
\usepackage[centertags]{amsmath}
\usepackage{amsbsy}
\usepackage{amsfonts}
\usepackage{amssymb}
\usepackage{graphicx}
\usepackage{bm}
\usepackage{cite}
\newcommand{\nua}[1]{\ensuremath{\rlap
           {\kern-2.5pt\ensuremath
           {\overset{\scriptscriptstyle(-)}{\phantom{\nu}}}}
           {\ensuremath{{\nu}_{#1}}}}}
\begin{document}

\title{Phenomenology of neutrino oscillations and mixing\thanks{
Talk presented by Marco Laveder at the XXXVII International Conference
of Theoretical Physics ``Matter to the deepest'',
Ustro\'n, 1--6 September 2013.
}}

\author{
Marco Laveder
\vspace{-0.3cm}
\address{Dipartimento di Fisica e Astronomia ``G. Galilei'', Universit\`a di Padova,
and\\
INFN, Sezione di Padova,
Via F. Marzolo 8, I--35131 Padova, Italy}
\\
and
\\
Carlo Giunti
\vspace{-0.3cm}
\address{INFN, Sezione di Torino, Via P. Giuria 1, I--10125 Torino, Italy}
}

\maketitle

\begin{abstract}
We review the
status of three-neutrino mixing and
the results of global analyses of short-baseline neutrino oscillation data
in 3+1 and 3+2
neutrino mixing schemes.
\end{abstract}


\section{Introduction}
\label{Introduction}

Neutrino oscillations have been measured with high accuracy in
solar, atmospheric and long-baseline
neutrino oscillation experiments.
Hence, we know that
neutrinos are massive and mixed particles
(see \cite{Giunti-Kim-2007}).
In this short review we discuss the status of the standard
three-neutrino mixing paradigm
(Section~\ref{Three})
and the indications in favor of the existence of additional sterile neutrinos
given by anomalies found in some short-baseline neutrino oscillation experiments
(Section~\ref{Beyond}).

\section{Three-Neutrino Mixing}
\label{Three}

Solar neutrino experiments
(Homestake,
Kamiokande,
GALLEX/GNO,
SAGE,
Super-Kamiokande,
SNO,
BOREXino)
measured $\nu_{e} \to \nu_{\mu}, \nu_{\tau}$
oscillations generated by the solar squared-mass difference
$
\Delta m^2_{\text{SOL}}
\simeq
7 \times 10^{-5} \, \text{eV}^2
$
and a mixing angle
$
\sin^2 \vartheta_{\text{SOL}}
\simeq
0.3
$.
The KamLAND experiment
confirmed these oscillations by observing the disappearance
of reactor $\bar\nu_{e}$ at an average distance of about 180 km.

Atmospheric neutrino experiments
(Kamiokande,
IMB,
Super-Kamiokande,
MACRO,
Soudan-2,
MINOS)
measured $\nu_{\mu}$ and $\bar\nu_{\mu}$
disappearance through oscillations generated by the atmospheric squared-mass difference
$
\Delta m^2_{\text{ATM}}
\simeq
2.3 \times 10^{-3} \, \text{eV}^2
$
and a mixing angle
$
\sin^2 \vartheta_{\text{ATM}}
\simeq
0.5
$.
The K2K and MINOS long-baseline experiments
confirmed these oscillations by observing the disappearance
of accelerator $\nu_{\mu}$
at distances of about 250 km and 730 km, respectively.

The two independent solar and atmospheric $\Delta{m}^2$'s
are nicely accommodated in the standard framework of
three-neutrino mixing
in which the three active flavor neutrinos
$\nu_{e}$,
$\nu_{\mu}$,
$\nu_{\tau}$
are superpositions of
three neutrinos
$\nu_{1}$,
$\nu_{2}$,
$\nu_{3}$
with masses
$m_{1}$,
$m_{2}$,
$m_{3}$:
$
\nu_{\alpha}
=
\sum_{k=1}^{3}
U_{\alpha k}
\nu_{k}
$,
for
$\alpha=e,\mu,\tau$.
The unitary mixing matrix can be written in the standard parameterization
in terms of
three mixing angles
$\vartheta_{12}$,
$\vartheta_{23}$,
$\vartheta_{13}$
and a CP-violating phase\footnote{
For simplicity,
we do not consider the two Majorana CP-violating phases which
contribute to neutrino mixing if massive neutrinos are Majorana particles,
because they do not affect neutrino oscillations
(see \cite{Giunti-Kim-2007}).
}
$\delta$:
\begin{equation}
U
=
\begin{pmatrix}
c_{12}
c_{13}
&
s_{12}
c_{13}
&
s_{13}
e^{-i\delta}
\\
-
s_{12}
c_{23}
-
c_{12}
s_{23}
s_{13}
e^{i\delta}
&
c_{12}
c_{23}
-
s_{12}
s_{23}
s_{13}
e^{i\delta}
&
s_{23}
c_{13}
\\
s_{12}
s_{23}
-
c_{12}
c_{23}
s_{13}
e^{i\delta}
&
-
c_{12}
s_{23}
-
s_{12}
c_{23}
s_{13}
e^{i\delta}
&
c_{23}
c_{13}
\end{pmatrix}
\,,
\label{U}
\end{equation}
where
$ c_{ab} \equiv \cos\vartheta_{ab} $
and
$ s_{ab} \equiv \sin\vartheta_{ab} $.
It is convenient to choose the numbers of the massive neutrinos
in order to have
\begin{equation}
\Delta{m}^{2}_{\text{SOL}}
=
\Delta{m}^{2}_{21}
,
\qquad
\Delta{m}^{2}_{\text{ATM}}
=
|\Delta{m}^{2}_{31}|
\simeq
|\Delta{m}^{2}_{32}|
,
\label{dm2}
\end{equation}
with
$\Delta{m}^{2}_{kj} = m_{k}^2 - m_{j}^2$.
Then,
there are two possible hierarchies for the neutrino masses:
the normal hierarchy
(NH)
with
$m_{1}<m_{2}<m_{3}$
and
the inverted hierarchy
(IH)
with
$m_{3}<m_{1}<m_{2}$.

With the conventions in Eqs.~(\ref{U}) and (\ref{dm2}),
we have
$\vartheta_{\text{SOL}}=\vartheta_{12}$
and
$\vartheta_{\text{ATM}}=\vartheta_{23}$.
Moreover,
the mixing angle
$\vartheta_{13}$
generates
$\nua{e}$ disappearance
and
$\nua{\mu}\to\nua{e}$ transitions
driven by
$\Delta{m}^{2}_{\text{ATM}}$,
which can be observed in long-baseline neutrino oscillation experiments.

In 2011 the T2K experiment reported the first indication of long-baseline
$\nu_{\mu}\to\nu_{e}$
transitions
\cite{1106.2822},
followed by the MINOS experiment
\cite{1108.0015}.
Recently,
the T2K Collaboration reported a convincing
$7.5\sigma$
observation of
$\nu_{\mu}\to\nu_{e}$
transitions
through the measurement of
28
$\nu_{e}$
events
with an expected background of
$4.64 \pm 0.53$
events
\cite{Malek-Lomonosov-2013}.

On the other hand,
the most precise measurement of the value of $\vartheta_{13}$
comes from the measurement of $\bar\nu_{e}$
disappearance in the Daya Bay
reactor experiment
\cite{1203.1669},
which has been confirmed by the data of the RENO
\cite{1204.0626}
and Double Chooz
\cite{1301.2948}
reactor experiments:
\begin{equation}
\sin^{2}2\vartheta_{13}
=
0.090 {}^{+0.008}_{-0.009}
\qquad
\text{\protect\cite{Jetter-NuFact-2013}}
\,.
\label{t13}
\end{equation}
Hence, we have a robust evidence of a non-zero value of $\vartheta_{13}$,
which is very important,
because the measured value of $\vartheta_{13}$
opens promising perspectives for the observation of CP violation in the lepton sector
and matter effects in long-baseline oscillation experiments,
which could allow to distinguish the normal and inverted neutrino mass spectra
(see \cite{1003.5800}).

As a result of all these observations of neutrino oscillations,
the mixing parameters can be determined with good precision
by a global fit of the data
\cite{1205.4018,1205.5254,1209.3023}.
The most recent result is NuFIT-v1.2
\cite{NuFIT}:
\begin{align}
\null & \null
\Delta{m}^2_{21}
=
7.45 {}^{+0.19}_{-0.16} \times 10^{-5} \, \text{eV}^2
\,, \quad
\sin^2\vartheta_{12}
=
0.306 {}^{+0.012}_{-0.012}
\,,
\label{p1}
\\
\null & \null
\Delta{m}^2_{31}
=
2.417 {}^{+0.013}_{-0.013} \times 10^{-3} \, \text{eV}^2
\,, \quad
\sin^2\vartheta_{23}
=
0.446 {}^{+0.007}_{-0.007}
\quad
\text{(NH)}
\,,
\label{p2}
\\
\null & \null
\Delta{m}^2_{32}
=
- 2.410 {}^{+0.062}_{-0.062} \times 10^{-3} \, \text{eV}^2
\,, \quad
\sin^2\vartheta_{23}
=
0.587 {}^{+0.032}_{-0.037}
\quad
\text{(IH)}
\,,
\label{p3}
\\
\null & \null
\sin^2\vartheta_{13}
=
0.0229 {}^{+0.0020}_{-0.0019}
\,.
\label{p4}
\end{align}
Hence,
the squared-mass differences are known with good precision:
about
$2.5\%$ for both
$\Delta{m}^2_{21}$
and
$|\Delta{m}^2_{31}|\simeq|\Delta{m}^2_{32}|$.
The mixing parameters
$\sin^2\vartheta_{12}$,
$\sin^2\vartheta_{13}$,
$\sin^2\vartheta_{23}$
are known, respectively,
with
$4\%$,
$9\%$,
$10\%$
precision.
Currently,
the most puzzling uncertainty is that of the mixing angle
$\vartheta_{23}$,
which is known to be close to the maximal mixing value of $\pi/4$,
but we do not know if it is smaller or larger.

We conclude this section noting a small tension
between reactor and accelerator measurements
of the $\vartheta_{13}$ angle.
It may be reconciled within the three-neutrino mixing scheme
by fitting the phase $\delta$ \cite{NuFIT}.
However, from an experimental point of view,
T2K shows an anomalous event vertex distribution of electron like events,
with the events concentrated near the border of the detector
\cite{Malek-Lomonosov-2013}.

\section{Beyond Three-Neutrino Mixing: Sterile Neutrinos}
\label{Beyond}

The completeness of the three-neutrino mixing paradigm has been challenged by
the following indications in favor of short-baseline neutrino oscillations,
which require the existence of at least one additional squared-mass difference,
$\Delta{m}^2_{\text{SBL}}$,
which is much larger than
$\Delta{m}^2_{\text{SOL}}$
and
$\Delta{m}^2_{\text{ATM}}$:
A)
The
LSND experiment,
in which a signal of short-baseline
$\bar\nu_{\mu}\to\bar\nu_{e}$
oscillations has been observed
with a statistical significance of about $3.8\sigma$
\cite{hep-ex/0104049}.
B)
The reactor antineutrino anomaly
\cite{1101.2755},
which is a $\sim2.8\sigma$ deficit of the rate of $\bar\nu_{e}$ observed in several
short-baseline reactor neutrino experiments
in comparison with that expected from a new calculation of
the reactor neutrino fluxes
\cite{1101.2663,1106.0687}.
C)
The Gallium neutrino anomaly
\cite{Laveder:2007zz},
consisting in a short-baseline disappearance of $\nu_{e}$
measured in the
Gallium radioactive source experiments
GALLEX
and
SAGE
with a statistical significance of about $2.9\sigma$.

In this review, we consider
3+1
\cite{hep-ph/9606411,hep-ph/9607372}
and
3+2
\cite{hep-ph/0305255}
neutrino mixing schemes
in which there are one or two additional massive neutrinos at the eV scale
and
the masses of the three standard massive neutrinos are much smaller.
Since from the LEP measurement of the invisible width of the $Z$ boson
we know that there are only three active neutrinos
(see \cite{Giunti-Kim-2007}),
in the flavor basis the additional massive neutrinos correspond to
sterile neutrinos
\cite{Pontecorvo:1968fh},
which do not have standard weak interactions.

In the 3+1 scheme,
the effective probability of
$\nua{\alpha}\to\nua{\beta}$
transitions in short-baseline experiments has the two-neutrino-like form
\begin{equation}
P_{\nua{\alpha}\to\nua{\beta}}
=
\delta_{\alpha\beta}
-
4 |U_{\alpha4}|^2 \left( \delta_{\alpha\beta} - |U_{\beta4}|^2 \right)
\sin^2\!\left( \dfrac{\Delta{m}^2_{41}L}{4E} \right)
\,,
\label{pab}
\end{equation}
where $U$ is the mixing matrix,
$L$ is the source-detector distance,
$E$ is the neutrino energy and
$\Delta{m}^2_{41} = m_{4}^2 - m_{1}^2 = \Delta{m}^2_{\text{SBL}} \sim 1 \, \text{eV}^2$.
The electron and muon neutrino and antineutrino appearance and disappearance
in short-baseline experiments
depend on
$|U_{e4}|^2$ and $|U_{\mu4}|^2$,
which
determine the amplitude
$\sin^22\vartheta_{e\mu} = 4 |U_{e4}|^2 |U_{\mu4}|^2$
of
$\nua{\mu}\to\nua{e}$
transitions,
the amplitude
$\sin^22\vartheta_{ee} = 4 |U_{e4}|^2 \left( 1 - |U_{e4}|^2 \right)$
of
$\nua{e}$
disappearance,
and
the amplitude
$\sin^22\vartheta_{\mu\mu} = 4 |U_{\mu4}|^2 \left( 1 - |U_{\mu4}|^2 \right)$
of
$\nua{\mu}$
disappearance.

Since the oscillation probabilities of neutrinos and antineutrinos are related by
a complex conjugation of the elements of the mixing matrix
(see \cite{Giunti-Kim-2007}),
the effective probabilities of short-baseline
$\nu_{\mu}\to\nu_{e}$ and $\bar\nu_{\mu}\to\bar\nu_{e}$
transitions are equal.
Hence,
the 3+1 scheme cannot explain a possible CP-violating difference of
$\nu_{\mu}\to\nu_{e}$ and $\bar\nu_{\mu}\to\bar\nu_{e}$
transitions in short-baseline experiments.
In order to allow this possibility,
one must consider a 3+2 scheme,
in which, there are four additional effective mixing parameters in short-baseline experiments:
$\Delta{m}^2_{51}$,
which is conventionally assumed $\geq\Delta{m}^2_{41}$,
$|U_{e5}|^2$, $|U_{\mu5}|^2$
and
$\eta = \text{arg}\left[U_{e4}^*U_{\mu4}U_{e5}U_{\mu5}^*\right]$.
Since this complex phase appears with different signs in
the effective 3+2 probabilities of short-baseline
$\nu_{\mu}\to\nu_{e}$ and $\bar\nu_{\mu}\to\bar\nu_{e}$
transitions, it can generate measurable CP violations.

\begin{table}[t]
\begin{center}
\begin{tabular}{c|cccc|cc}
					&3+1							&3+1							&3+1							&3+1							&3+2							&3+2							\\
					&LOW							&HIG							&noMB							&noLSND							&LOW							&HIG							\\
\hline
$\chi^{2}_{\text{min}}$			&291.7		&261.8		&236.1		&278.4		&284.4		&256.4		\\
NDF					&256		&250		&218		&252		&252		&246		\\
GoF					& 6\%		&29\%		&19\%		&12\%		& 8\%		&31\%		\\
\hline
$(\chi^{2}_{\text{min}})_{\text{APP}}$	&99.3		&77.0		&50.9		&91.8		&87.7		&69.8		\\
$(\chi^{2}_{\text{min}})_{\text{DIS}}$	&180.1		&180.1		&180.1		&180.1		&179.1		&179.1		\\
	$\Delta\chi^{2}_{\text{PG}}$	&12.7		&4.8		&5.1		&6.4		&17.7		&7.5		\\
	$\text{NDF}_{\text{PG}}$	&2		&2		&2		&2		&4		&4		\\
	$\text{GoF}_{\text{PG}}$	&0.2\%	& 9\%	& 8\%	& 4\%	&0.1\%	&11\%	\\
\hline
$\Delta\chi^{2}_{\text{NO}}$		&$47.5$		&$46.2$		&$47.1$		&$8.3$		&$54.8$		&$51.6$		\\
	$\text{NDF}_{\text{NO}}$	&$3$		&$3$		&$3$		&$3$		&$7$		&$7$		\\
	$n\sigma_{\text{NO}}$		&$6.3\sigma$	&$6.2\sigma$	&$6.3\sigma$	&$2.1\sigma$	&$6.0\sigma$	&$5.8\sigma$	\\
\end{tabular}
\end{center}
\caption{ \label{tab:chi}
\footnotesize
Results of the fit of short-baseline data
\cite{1308.5288}
taking into account all MiniBooNE data (LOW),
only the MiniBooNE data above 475 MeV (HIG),
without MiniBooNE data (noMB)
and without LSND data (noLSND)
in the
3+1 and
3+2 schemes.
The first three lines give
the minimum $\chi^{2}$ ($\chi^{2}_{\text{min}}$),
the number of degrees of freedom (NDF) and
the goodness-of-fit (GoF).
The following five lines give the quantities
relevant for the appearance-disappearance (APP-DIS) parameter goodness-of-fit (PG).
The last three lines give
the difference between the $\chi^{2}$ without short-baseline oscillations and $\chi^{2}_{\text{min}}$
($\Delta\chi^{2}_{\text{NO}}$),
the corresponding difference of number of degrees of freedom ($\text{NDF}_{\text{NO}}$)
and the resulting
number of $\sigma$'s ($n\sigma_{\text{NO}}$) for which the absence of oscillations is disfavored.
}
\end{table}

Global fits of short-baseline neutrino oscillation data have been presented recently in
Refs.~\cite{1303.3011,1308.5288}.
In the following we summarize the results of the analysis of short-baseline data
in the 3+1 and 3+2 schemes
presented in
Ref.~\cite{1308.5288}.
The statistical results are listed in Table~\ref{tab:chi}.
In the LOW fits
all the MiniBooNE data are considered,
including the anomalous low-energy bins,
which are omitted in the HIG fits.
There is also a 3+1-noMB fit without MiniBooNE data
and
a 3+1-noLSND fit without LSND data.

From Tab.~\ref{tab:chi},
one can see that in all fits which include the LSND data
the absence of short-baseline oscillations
is disfavored by about $6\sigma$,
because the improvement of the $\chi^2$ with short-baseline oscillations
is much larger than the number of oscillation parameters.

In both the
3+1 and 3+2 schemes,
the goodness-of-fit in the LOW analysis is significantly worse than that in the HIG analysis
and the appearance-disappearance parameter goodness-of-fit is much worse.
This result confirms the fact that the MiniBooNE low-energy anomaly
is incompatible with neutrino oscillations,
because it would require a small value of $\Delta{m}^2_{41}$
and a large value of $\sin^22\vartheta_{e\mu}$
which are excluded by the data of other experiments
(see Ref.~\cite{1308.5288} for further details).
Note that the appearance-disappearance tension
in the 3+2-LOW fit is even worse than that in the 3+1-LOW fit,
since the $\Delta\chi^{2}_{\text{PG}}$ is so much larger that it cannot be compensated
by the additional degrees of freedom.
Therefore,
we think that it is very likely that the MiniBooNE low-energy anomaly
has an explanation which is different from neutrino oscillations
and the HIG fits are more reliable than the LOW fits.

The 3+2 mixing scheme,
was considered to be interesting in 2010
when the MiniBooNE neutrino
\cite{0812.2243}
and antineutrino
\cite{1007.1150}
data showed a CP-violating tension.
Unfortunately,
this tension reduced considerably in the final MiniBooNE data
\cite{1303.2588}
and from Tab.~\ref{tab:chi}
one can see that there is little improvement of the 3+2-HIG fit
with respect to the 3+1-HIG fit,
in spite of the four additional parameters and the additional possibility of CP violation.
Moreover,
since the p-value obtained by restricting the 3+2 scheme to 3+1
disfavors the 3+1 scheme only at
$1.2\sigma$
\cite{1308.5288},
we think that considering the larger complexity of the 3+2 scheme
is not justified by the data\footnote{
See however the somewhat different conclusions reached in Ref.~\cite{1303.3011}.
}.

\begin{figure*}[t]
\null
\hfill
\includegraphics*[width=0.49\linewidth]{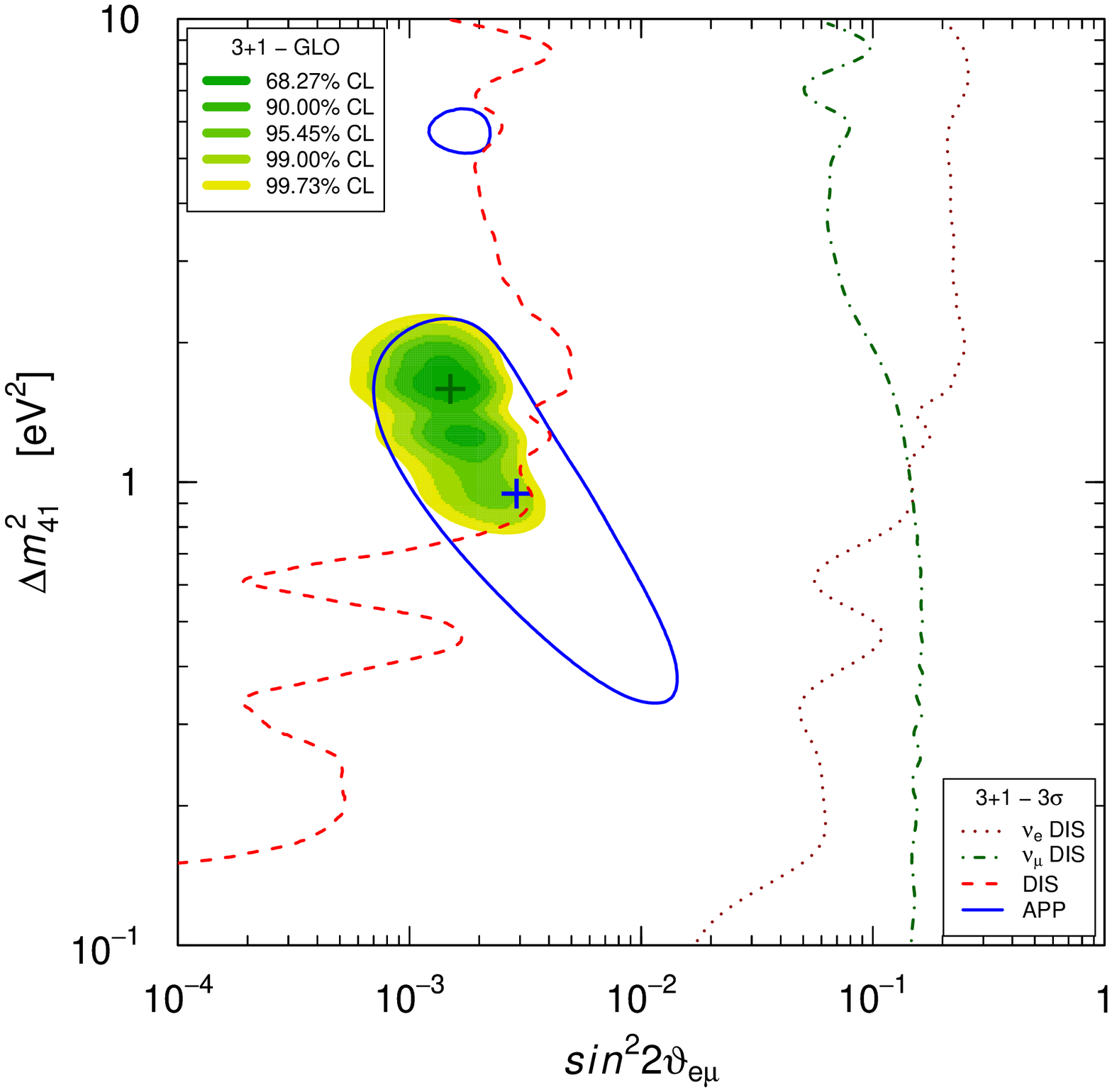}
\hfill
\includegraphics*[width=0.49\linewidth]{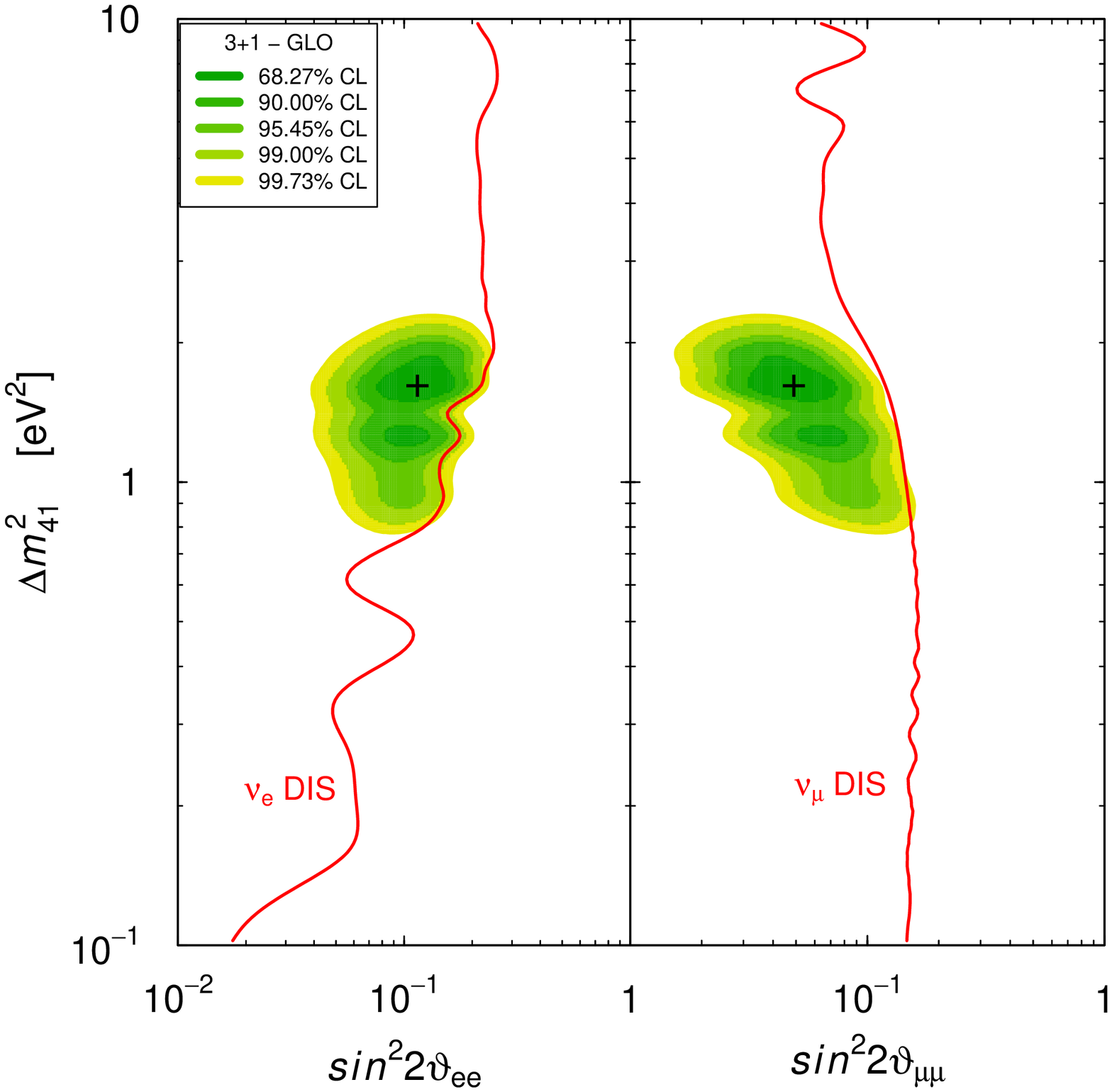}
\hfill
\null
\caption{ \label{fig:glo}
\footnotesize
Allowed regions in the
$\sin^{2}2\vartheta_{e\mu}$--$\Delta{m}^{2}_{41}$,
$\sin^{2}2\vartheta_{ee}$--$\Delta{m}^{2}_{41}$
and
$\sin^{2}2\vartheta_{\mu\mu}$--$\Delta{m}^{2}_{41}$
planes
obtained in the global (GLO) 3+1-HIG fit
\cite{1308.5288}
of short-baseline neutrino oscillation data
compared with the $3\sigma$ allowed regions
obtained from
$\protect\nua{\mu}\to\protect\nua{e}$
short-baseline appearance data (APP)
and the $3\sigma$ constraints obtained from
$\protect\nua{e}$
short-baseline disappearance data ($\nu_{e}$ DIS),
$\protect\nua{\mu}$
short-baseline disappearance data ($\nu_{\mu}$ DIS)
and the
combined short-baseline disappearance data (DIS).
The best-fit points of the GLO and APP fits are indicated by crosses.
}
\end{figure*}

Figure~\ref{fig:glo}
shows the allowed regions in the
$\sin^{2}2\vartheta_{e\mu}$--$\Delta{m}^{2}_{41}$,
$\sin^{2}2\vartheta_{ee}$--$\Delta{m}^{2}_{41}$ and
$\sin^{2}2\vartheta_{\mu\mu}$--$\Delta{m}^{2}_{41}$
planes
obtained in the 3+1-HIG fit of Ref.~\cite{1308.5288}.
These regions are relevant, respectively, for
$\nua{\mu}\to\nua{e}$ appearance,
$\nua{e}$ disappearance and
$\nua{\mu}$ disappearance
searches.
Figure~\ref{fig:glo}
shows also the region allowed by $\nua{\mu}\to\nua{e}$ appearance data
and
the constraints from
$\nua{e}$ disappearance and
$\nua{\mu}$ disappearance data.
One can see that the combined disappearance constraint
in the $\sin^{2}2\vartheta_{e\mu}$--$\Delta{m}^{2}_{41}$ plane
excludes a large part of the region allowed by $\nua{\mu}\to\nua{e}$ appearance data,
leading to the well-known
appearance-disappearance tension
quantified by the parameter goodness-of-fit in Tab.~\ref{tab:chi}.

It is interesting to investigate what is the
impact of the MiniBooNE experiment
on the global analysis of short-baseline neutrino oscillation data.
With this aim,
the authors of Ref.~\cite{1308.5288}
performed two additional 3+1 fits:
a 3+1-noMB fit without MiniBooNE data
and
a 3+1-noLSND fit without LSND data.
From Tab.~\ref{tab:chi}
one can see that the results of the
3+1-noMB fit are similar to those of the
3+1-HIG fit
and the exclusion of the case of no-oscillations remains at the level of $6\sigma$.
On the other hand,
in the 3+1-noLSND fit,
without LSND data,
the exclusion of the case of no-oscillations drops dramatically to
$2.1\sigma$.
In fact,
in this case
the main indication in favor of short-baseline oscillations
is given by the reactor
and
Gallium
anomalies
which have a similar statistical significance.
Therefore,
it is clear that the LSND experiment is still crucial for the indication in favor of short-baseline
$\bar\nu_{\mu}\to\bar\nu_{e}$
transitions
and the MiniBooNE experiment has been rather inconclusive.

\section{Conclusions}
\label{Conclusions}

The current status of our knowledge of three-neutrino mixing is very satisfactory
after the recent determination of the smallest mixing angle $\vartheta_{13}$:
the two squared-mass differences and the three mixing angles are known with good precision.
Future experiments must determine
if $\vartheta_{23}$ is smaller or larger than $\pi/4$,
the value of the Dirac CP-violating phase in the mixing matrix,
the mass hierarchy
and the absolute scale of neutrino masses.
It is also very important to find if neutrinos are Majorana particles
and in that case what are the values of the Majorana CP-violating phases.

Anomalies which cannot be explained in the framework of three-neutrino mixing
and require the existence of sterile neutrinos
have been observed by some short-baseline neutrino oscillation experiments.
The results of the global fit of
short-baseline neutrino oscillation data presented
in Ref.~\cite{1308.5288}
show that the data can be explained by 3+1 neutrino mixing
and this simplest scheme beyond three-neutrino mixing
cannot be rejected in favor of
the more complex 3+2 scheme.
The low-energy MiniBooNE anomaly cannot be explained by neutrino oscillations
in any of these schemes.
Moreover,
the crucial indication
in favor of short-baseline
$\bar\nu_{\mu}\to\bar\nu_{e}$
appearance is still given by the old LSND data
and the MiniBooNE experiment has been inconclusive.
Hence new better experiments are needed in order to
check this signal.



\end{document}